\documentclass [preprint, aps, amssymb, amsmath, showpacs]{revtex4}

\usepackage{graphicx}

\newcommand{\pa}{phenylene-acetylene }
\newcommand{\Pa}{Phenylene-acetylene }

\begin{document}

\preprint{Submitted to Chemical Physics Letters}

\title{A joint theoretical and experimental study of phenylene-acetylene
molecular wires}

\author{R. J. Magyar and S. Tretiak}\email{serg@cnls.lanl.gov}
\affiliation{Theoretical Division and Center for Nonlinear Studies,
Los Alamos National Laboratory,
Los Alamos, NM 87545}

\author{Y.  Gao,  H.-L.  Wang  and  A. P. Shreve}  \affiliation{Bioscience
Division, Los Alamos National Laboratory, Los Alamos, NM 87545}

\begin{abstract}
The   excited   state  electronic   structure   of  $\pi$   conjugated
phenylene-acetylene  oligomers is calculated  using time-dependent
density  functional   theory  (TD-DFT)  approaches.   The  theoretical
fluorescence  spectra are  analyzed  in terms  of Frank-Condon  active
nuclear   normal    modes   and    shown   to   compare    well   with
experiment.  Theoretical  and  experimental  results for  the  optical
absorption and  emission spectra of these molecules  indicate that the
conjugation  length  can be  significantly  reduced by  conformational
rotations  about  the triple-bonded  carbon  links.  This has  serious
implications     on      the     electronic     functionalities     of
polyphenylene-acetylene based  molecular wires and  their possible use
as charge/energy conduits in nano-assemblies.

\end{abstract}


\date{\today}

\maketitle

\section{Introduction}
\label{sec.introduction}

In  nature, complex  nano-assemblies  of functional  units combine  to
perform required  complicated tasks. One  example is the  assembly for
harvesting             light             in             photosynthesis
\cite{PulleritsT:Pholpc,SundstromV:PholRd}.    There  is  considerable
practical  interest  in  imitating   and  manipulating  this  sort  of
functionality  on a  nano-scale. In  order to  do so,  theoretical and
experimental understanding  of the constituents is  required.  In this
paper,  we  consider  a  functional  component  of  a  possible  light
harvesting  assembly.   One  functional  unit of  a  light  harvesting
assembly  might  be  a  self-assembled multilayer  with  the  specific
function of  transferring holes away from  light receptors. Conjugated
polymers  may be  good candidates  for this  purpose because  of their
semiconductor-like  properties, plasticity,  and potential low cost of
fabrication \cite{FriendRH:Elecp}. These  materials have already found
uses in  a wide range  of applications such as  light-emitting diodes,
lasers,         sensors         and         molecular         switches
\cite{SchmitzC:Polldb,HideF:Sempnc,YangJS:Porspf,Brownar:Loggmf}.

To  perform  such functions,  the  molecule  should  be rigid  in  one
dimension  to  provide   required  spatial  conduits.   Moreover,  the
hole-electron separation and  charge conduction processes require long
conjugation lengths  and flexible  excited state structure  to provide
sufficient  freedom  to tune  energetics  synthetically.  We  consider
oligomers of \pa,  shown on the top panel  of Fig.~\ref{fig:pa}, which
retain one-dimensional  rigidity like molecular  wires.  However, this
polymer  can  be  easily  twisted  between  benzene  rings  about  the
triple-bonded carbon links. This geometry change is expected to reduce
the  conjugation length  and may  adversely affect  applications which
rely on the de-localization of the charge carriers. In this article we
examine   \pa  oligomers  from   both  theoretical   and  experimental
standpoints  in  order  to  analyze  their  excited  state  electronic
structure  and the  effects of  conformational rotations  on effective
conjugation   and   electronic   de-localization  lengths   of   these
materials. In  particular, we utilized the  natural transition orbital
decomposition  \cite{MartinRL:Natto}  to   infer  the  extent  of  the
underlying electronic localization.

To calculate electronic  structure we use a blend  of quantum chemical
methods  including semi-empirical  (Austin Model  1  (AM1)) approaches
\cite{DewarMJS:AM1ngp}     for     ground     and    excited     state
\cite{TretiakS:Condpc}  geometry   optimizations,  and  time-dependent
density  functional  theory  (TD-DFT) \cite{Rungee:Dentft,casida}  and
ZINDO  \cite{RidleyJ:intndo} methods  for excited  state computations.
There have  been many theoretical calculations  on conjugated polymers
using  various semi-empirical  and first  principle approaches.  It is
well understood  that, in general,  both semi-empirical (such  as AM1)
and DFT  methods provide reasonable  geometries and ZINDO  can deliver
good  UV-visible     spectra    for    such     molecular    systems
\cite{BredasJL:Excesc,TretiakS:Denmas}.   Even  more  accurate  TD-DFT
approaches are able to  tackle complicated electronic features such as
exciton                         binding                         energy
\cite{RuiniA:Solsee,vanderHorstJW:Abice,RohlfingM:Optecp}  and  double
excitations \cite{MaitraNT:Memtdf}. Furthermore,  based on the results
of the  calculations, absorption an  emission spectra can  be modeled,
for  example,  for  polyenes, oligoarylenes,  phenylenevinylenes,  and
polyfluorenes              \cite{KarabunarlievS:RigFae,TretiakS:Condpc,
FrancoI:Elerpp}.  Existing  theoretical studies of  \pa include TD-DFT
investigation    of    the    excited    states   of    the    monomer
\cite{Serrano-AndresL:Theesa,AmatatsuY:Abisp}  and several  studies of
dendrimer      with      \pa       being      the      basic      unit
\cite{ThompsonAL:Varecp,TretiakS:Loceep}.   Additionally, experimental
absorption  and  fluorescence  spectra  have  been  reported  for  \pa
monomers,    dimers,    and    trimers    linked   to    Pt    centers
\cite{McLeanDG:Speakb}.

Details   of    our   computational   approach    are   presented   in
Section~\ref{sec.methods}.   In Section  \ref{sec.results}  we analyze
computational  results and  compare them  to experiment.   Finally, we
discuss  the  trends  that   emerge  and  summarize  our  findings  in
Section~\ref{sec.conclusion}.

\section{Computational Methodology}
\label{sec.methods}

We focus  on oligomers of  two to ten  \pa repeat-units (top  panel of
Fig.~\ref{fig:pa}).  The two unit case is the small molecule limit and
may be compared  to more expensive {\em ab  initio} calculations which
are  only  possible  for  small  systems.   In  such  small  molecules
photo-excitations  are  confined  by   molecular  ends  and  study  of
increasingly  longer chains is  necessary to  understand the  onset of
de-localized excitations that polymers  exhibit. The ten unit chain is
a sufficiently  long oligomer  to reasonably approximate  the infinite
polymer limit \cite{BredasJL:Excesc,TretiakS:Denmas}.

Ground state  optimal geometries of  \pa oligomers have  been obtained
using  the Gaussian  98  \cite{g98} software  package.  The  molecular
geometries   in  the  gas   phase  are   fully  optimized   using  the
semi-empirical AM1 method. The AM1 model has been parametrized to give
accurate geometries  for organic molecules and is  expected to provide
reliable geometries \cite{DewarMJS:AM1ngp}.  To model the fluorescence
spectra, the excited state geometries are needed. We used the
excited-state molecular dynamics (ESMD)
computational  package  \cite{TretiakS:Condpc}  to optimize  molecular
geometries for the lowest excited state at time-dependent Hartree-Fock
(TD-HF) level and AM1 model.   This approach allows treatment of large
molecular systems at modest  numerical cost and previously resulted in
reasonable  fluorescence line-shapes  of several  conjugated molecular
systems  \cite{TretiakS:Condpc,FrancoI:Elerpp}.  In  principle,  it is
also possible  to use DFT  or Hartree-Fock {\em ab  initio} approaches
for ground  state optimization purposes. However,  such approaches may
be  problematic  for  the  excited  state  optimization  due  to  high
numerical  cost.  Instead,  we treat  molecular geometries  within the
same approach: HF/AM1 and TD-HF/AM1  levels for the ground and excited
states, respectively. Additionally, by using the AM1, we eliminate any
errors  in  the  geometry  that  might  come  from  using  approximate
functionals and  limited basis sets in density  functional theory. For
example,  it   was  found  for  polyacetylene  that   a  fraction  of
exact-exchange  must be mixed  with a  semi-local exchange-correlation
functional  for the  theory to  reproduce the  bond-length alternation
accurately \cite{ChoiCH:Theeec}.

For obtained  geometries, we next calculate  the excited-state triplet
and singlet  manifolds using  TD-DFT approach which  is known to  be a
reliable  but  computationally affordable  \emph{ab  initio} tool  for
excited state  treatment. We perform  all our calculations in  the gas
phase for simplicity,  but we expect that the  results will not change
much when these oligomers are placed in non-polar solvents. We use the
B3LYP functional combined  with the 6-31G basis set  as implemented in
the Gaussian 98  package \cite{g98}.  The 6-31G basis  set is known to
be  an efficient  blend  of  accuracy and  manageable  size for  large
conjugated  molecules  \cite{MasunovAM:Pretap}.   We  do  not  include
diffuse  functions  as we  expect  the  relevant  excited states  have
support only  along the backbone of  the polymer for  long chains. For
ground-state  properties, DFT  provides  a formally  exact scheme  for
solving  the   many-body  problem  \cite{Hohenbergp:Inheg},   but,  in
practice, the functionals used are approximated in a manner convenient
for calculations.   The B3LYP functional  \cite{Beckead:Anmo} combines
semi-local  exchange-correlation  with  non-local exact-exchange.   By
construction,  this  functional   handles  a  fraction  of  long-range
exchange   exactly  but  fails   to  capture   long-range  correlation
effects. Time-dependent  density functional theory is  an extension of
density   functional  theory  in   which  many-body   excitations  are
associated   with   the   poles   of  the   exact   density   response
\cite{Rungee:Dentft,casida}. TD-DFT using  B3LYP inputs has been shown
to  be accurate  for  many molecular  systems  and is  computationally
affordable.   In particular,  Ref.  \cite{YuJSK:Timdfs}  suggests that
B3LYP is the optimal  functional to use for excited-state calculations
on PPV-type polymers.

In order to characterize calculated  excited states and to address the
electronic  localization, we performed  a transition  orbital analysis
\cite{MartinRL:Natto} based on  the computed transition densities from
the  TD-DFT  calculations.   This  analysis offers  the  most  compact
representation of a given transition density in terms of its expansion
in  single-particle  transitions.    The  transition  orbital  picture
provides an important  information on the multi-configurational nature
of   a  given   excited  state,   and  gives   a   real-space  orbital
representation as to where  the photo-excited hole and electron reside,
which is useful to illustrate the excitonic localization phenomena.

Finally, we  calculate the  line-shapes of fluorescence  spectra using
the ground and excited state optimal geometries and vibrational normal
modes of the ground state. This  can be readily done within the Condon
approximation  for  displaced  multidimensional  harmonic  oscillators
\cite{Myersab:Excgcf,KarabunarlievS:RigFae}.   The vibrational overlap
integrals     $|\langle      0     |     \nu_n\rangle      |^{2}     =
\frac{e^{-S_n}S_n^\nu}{\nu!}$, Franck-Condon   factors,   govern   the
probability of emission from  transition between the vibrational level
0  in the  lowest excited  state and  a vibrational  level $n$  in the
ground state.  These quantities,  in turn, depend on the dimensionless
displacements $\Delta_n$ of each  normal mode with Huang-Rhys factors,
$S_n= \Delta_n^{2}/2$.   The fluorescence band shape as  a function of
the  frequency $\omega$  is determined  by the  imaginary part  of the
polarizability \cite{Myersab:Excgcf,KarabunarlievS:RigFae}
\begin{equation}\label{eq:fluor}
\alpha(\omega) = \textrm{Im}\left\{\mu^{2}\sum_{\nu_1}\cdots\sum_{\nu_{3N-6}}
\frac{\Pi_{n=1}^{3N-6}\langle 0 | \nu_{n}\rangle^{2}}
     {\Omega^{(0)} - \sum_{n=1}^{3N-6}\nu_n\omega_n - \omega -
       i\Gamma}\right\},
\end{equation}
where  $\mu$ is the  electronic transition  dipole moment  between the
excited and  the ground-state, $\Omega^{(0)}$ is  the associated $0-0$
transition energy,  the $\omega_n$'s are  the vibrational frequencies,
the $\nu_n$'s  are the quanta  of the participating normal  modes, and
$\Gamma$ is  an empirical parameter setting  the spectral line-widths.
We choose the  line-width to be either 0.2 eV or  0.02 eV.  The former
produces plots which agree  well with the experimentally observed line
widths, whereas the latter  allows greater resolution and the analysis
in terms of the contributing vibrational states.

\section{Results and Discussions}
\label{sec.results}

\begin{figure}[ht]
  \centering
  \includegraphics[width=0.9\textwidth]{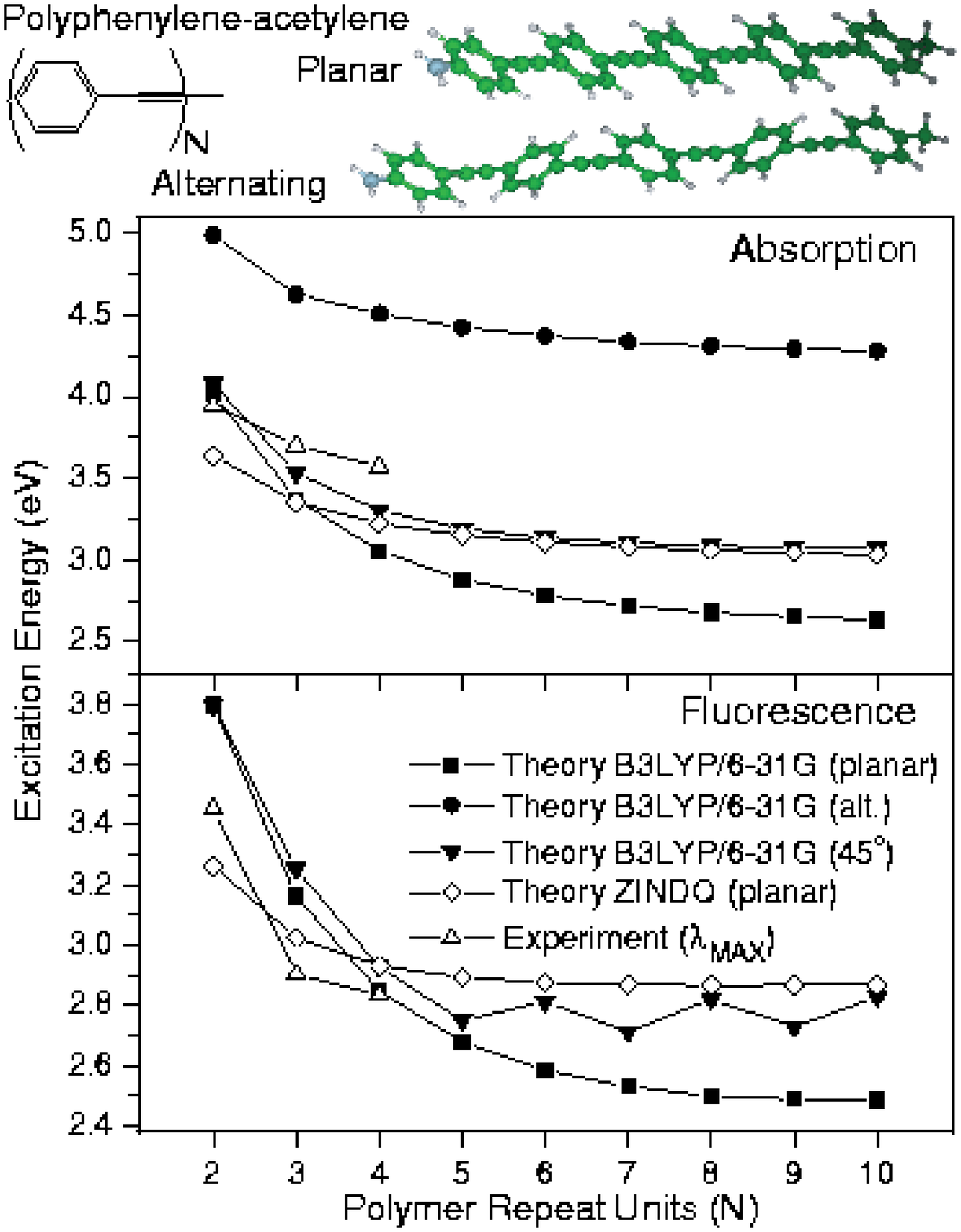}
 \vspace{-0.4in}
\caption{\label{fig:pa}
Top: molecular structure of \pa and schematic structures of the planar
and  alternating  geometries  respectively.   Middle: scaling  of  the
energy  of  the  singlet  excited  state  which  dominates  uv-visible
absorption  as a  function  of  the number  of  polymer repeat  units.
Theoretical values correspond to the vertical excitation and emission,
and  experimental values  are taken  for the  absorption  and emission
maxima.  Bottom: Same as above but for fluorescence.}
\end{figure}

The chemical  structure of \pa  is shown on  Fig.~(\ref{fig:pa}).  Our
specimen is  terminated at one end  by a methyl group  ($CH_3$) and at
the other end by an amine  group ($NH_2$) which may either be used for
binding  the molecule  to to  a  substrate or  for self-assembly  into
structures of  interest at the  air-water interface (Langmuir-Blodgett
method).   Oligomers  with  N=2,3,   and  4  repeat  units  have  been
synthesized  and spectroscopically  characterized.  The  \pa oligomers
were synthesized by  cross-coupling of the appropriate phenylacetylene
compounds  with   4-amino-1-iodobenzene  using  $Pd(PPh_3)_4$/CuI  as
catalyst. The  molecular building block (\Pa) is  prepared by coupling
4-dodecane-ethynylbenzene  with 4-(trimethylsilylethynyl) iodobenzene.
The  solution  UV-vis and  fluorescence  spectra  were measured  using
methylene chloride as the solvent.

We calculate the threshold for rotation about the triple bond is $\sim
0.05$ eV per triple bond, and we expect that at room temperature where
$k_BT\approx  0.025$ eV,  an ensemble  of geometries  will  be allowed
given  the uncertainty of  solvent effects  and molecular  packing. In
order to estimate how these geometry changes will affect the spectrum,
we sample  two extreme geometries (Fig.  \ref{fig:pa}), the completely
planar configuration  where all benzene  rings lie in the  same plane,
and an  alternating configuration where adjacent benzene  rings are at
right angles  to each other.  Both geometries are local  minima within
the  AM1 approach,  the planar  case being  the global  minimum.  Even
though in  total ground-state energies  the two configurations  do not
differ  drastically,  their  excited-states properties  do.  Torsional
potentials for another type of conjugated polymer have been calculated
in detail  \cite{KarpfenA:Sintpc}.  Here  we choose only  to calculate
the  extreme limits and  some intermediates  arguing that  the general
behavior and  trends will  be thereby evident  and the details  of the
torsional potential are not critical to the current study.

In  Fig~\ref{fig:pa}, we  plot the  dependence  of the  energy of  the
uv-visible  active singlet  state as  a function  of chain  length for
absorption and  fluorescence.  Absorption results are  presented for a
planar, a  completely alternating, and an  intermediate geometry where
every  other  benzene   is  rotated  about  the  triple   bond  by  45
degrees. The experimental results  are the absorption and fluorescence
maxima in a non-polar solvent,  methylene chloride.  Except for in the
alternating  geometry,   the  lowest  $S_1$  state   has  the  largest
oscillator  strength. We  see  that the  experiment  lies between  the
intermediate and completely alternating  results. The long chain limit
is reached by approximately four to six repeat units for planar chain,
which implies the extent of the conjugation length and the size of the
photo-generated exciton. This value is  a strong function of the exact
orbital-dependent  exchange   (Fock-like  exchange)  present   in  the
functional   \cite{kirill}.   For   example,  the   ZINDO   (100\%  of
exact-exchange)    curve   saturates    faster    to   the    constant
long-chain-limit  than B3LYP (20\%  of exact-exchange)  results. Since
the alternating  geometry breaks some  of the conjugation,  we observe
several  low-lying  dark singlets  energetically  below the  optically
active  \emph{band-gap}  state.   In  this  case, the  energy  of  the
\emph{band-gap} state has  little size-dependence and is significantly
blue-shifted compared  to that for planar geometry.   As expected, the
intermediate geometry  values lie between  these two extremes  and are
accidentally very  close to the  ZINDO results.  In order  to estimate
the  effective   bond  angle  (within   TD-B3LYP/6-31G  approach),  we
gradually  rotated  the  benzene  rings  about the  triple  bonds  and
calculated  the  singlet  excited  states  for  N=2,  3,  and  4  unit
oligomers. For most  angles, the lowest singlet state  can be directly
associated with the lowest singlet  state for the planar geometry.  At
some critical  angle, this  breaks down and  the lowest singlet  is no
longer optically  active. For 2-units,  the singlet and  experiment do
not  overlap  for  any angle.   For  3-units,  the  overlap is  at  43
degrees. The  effective rotation  angle for 4-unit  chains is  over 63
degrees.  In  the long chain limit,  we expect the  effective angle to
increase but saturate.

Geometric effects  are important for  fluorescence as well.   Here the
excited state geometries remain close to planar with a steep torsional
potential    as    has    been    reported    for    other    polymers
\cite{TretiakS:Condpc,FrancoI:Elerpp}.   The  bond-length  alternation
parameter reduces  in the  middle of the  molecule which  indicates an
excitonic  self-trapping  process.  Because  of  the  tendency of  the
excited state  to planarize the molecule, the  alternating geometry is
unlikely,  so we  simulate  what one  might  expect to  become of  the
alternating geometry by taking  the relaxed excited state geometry and
rotating  each benzene  by  45  degrees about  the  triple bond  while
leaving  two  to  three benzene  rings  in  the  center of  the  chain
coplanar. We see that  the fluorescence from the simulated alternating
geometry differs  only slightly from the planar  geometry (pointing to
the  short extent  of the  self-trapped  exciton). The  curve for  the
simulated case is not smooth because  of the scheme we use to simulate
the geometry. For  even numbered chains we have  two coplanar benzenes
in the center of the chain, but for odd numbered chains, we have three
coplanar benzenes  and a slightly  greater de-localization. Therefore,
the results  for the first excited  singlets of odd  chain lengths are
systematically slightly lower.

\begin{figure}[ht]
  \centering
  \includegraphics[width=0.9\textwidth]{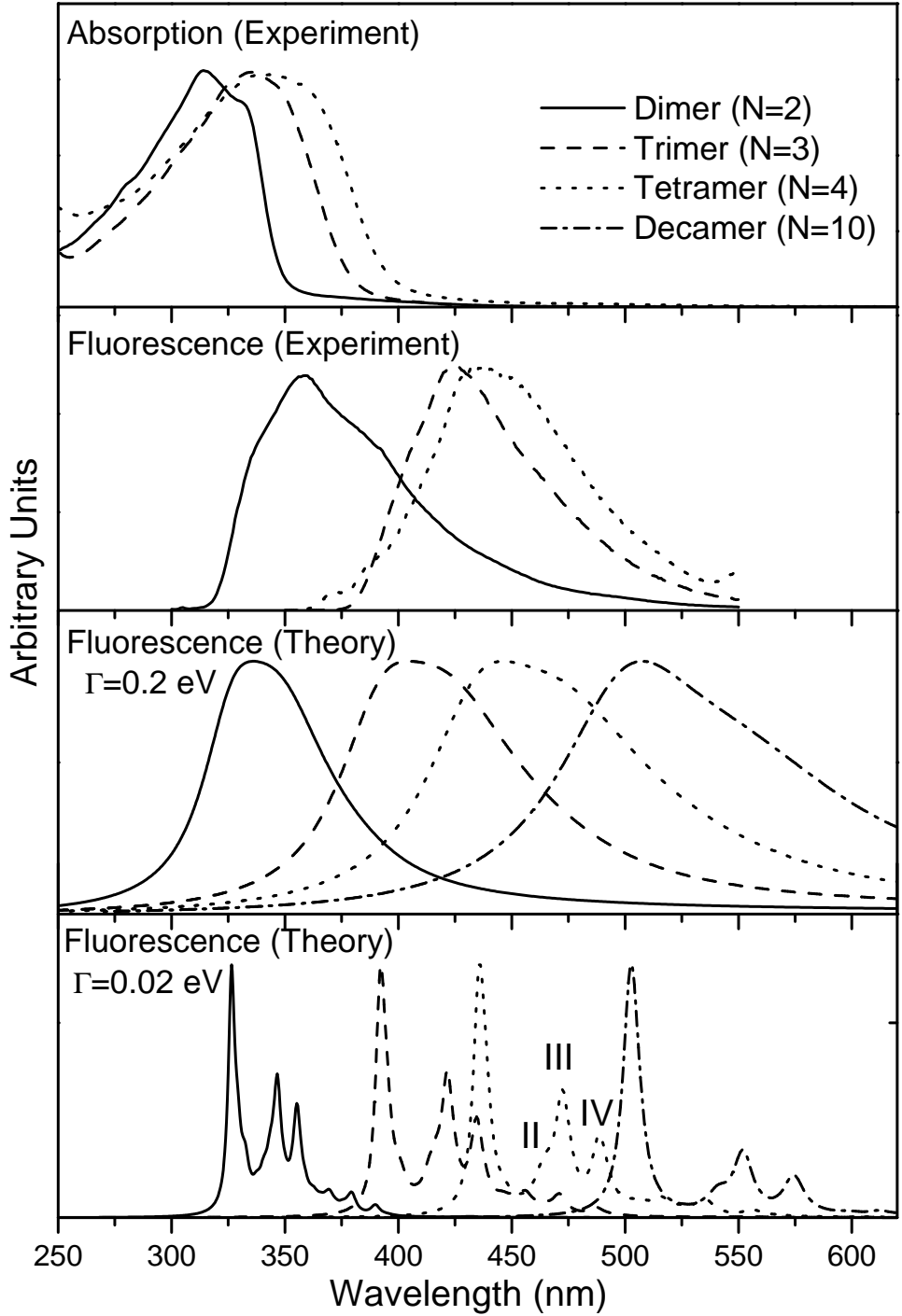}
 \vspace{-0.2in}
\caption{\label{fig:afplot}  Absorption and  fluorescence  line shapes
for  various  length \pa  oligomers;  experiment  versus
theory.  The  top  two  panels  are the  experimental  absorption  and
fluorescence.   The bottom  two  panels are  the theoretical  emission
profiles calculated with broad (0.2  eV) and narrow (0.02 eV) spectral
line width.  Frank-Condon active vibrational normal  modes (II-IV) are
shown in Fig.~\ref{fig:modes}. }
\end{figure}

\begin{figure}[ht]
  \centering
 \includegraphics[width=0.9\textwidth]{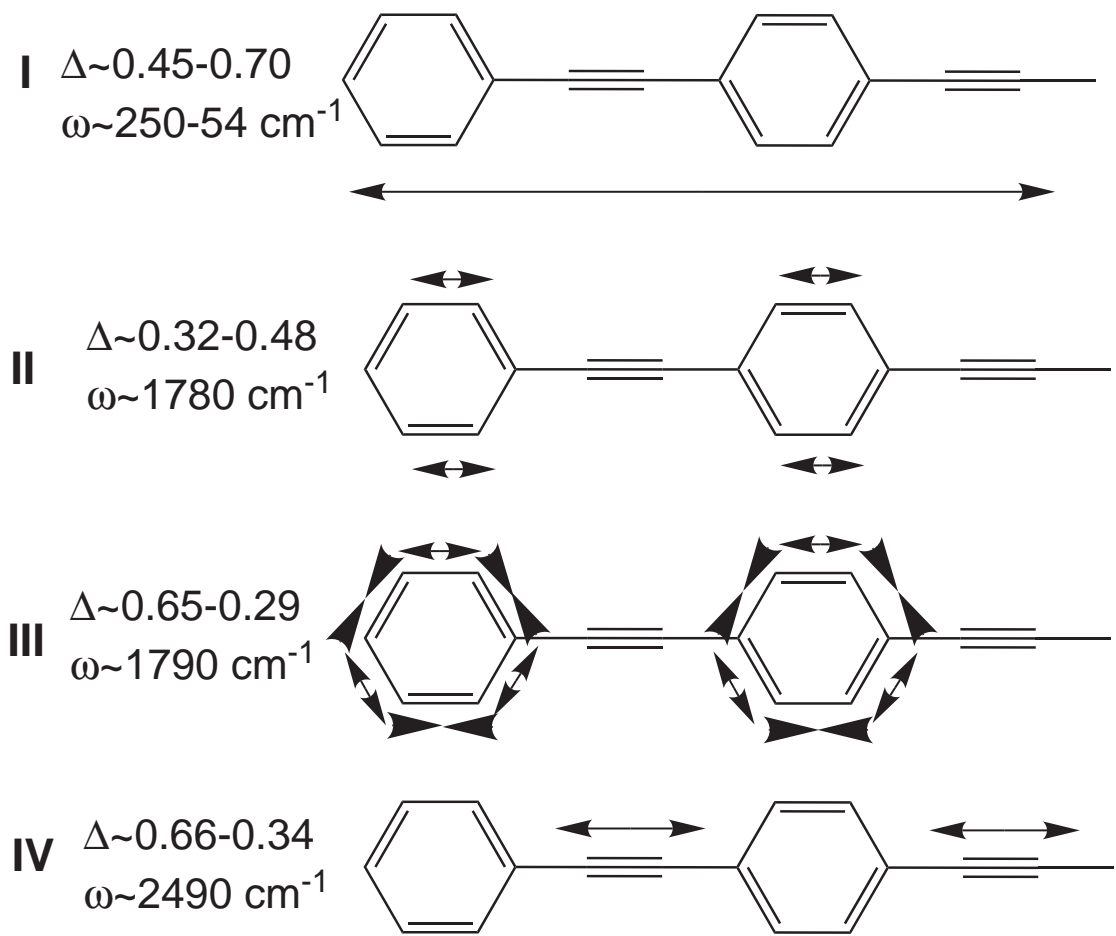}
 \vspace{0.2in}
\caption{\label{fig:modes} The  dominant normal modes  contributing to
the vibrational  structure of the  fluorescence spectrum for  \pa. The
top  diagram is  a  low  energy stretching  mode  contributing to  the
spectral   broadening.  The   next  two   diagrams   show  vibrations
contributing  to the II-III  peak, which  becomes resolved  as  the polymer
length increases;  these are oscillations  in the length of  the bonds
within the benzene rings. The  final mode is a high energy oscillation
in the length of the triple bond, which contributes to the IV peak. }
\end{figure}

Using Eq.~(\ref{eq:fluor}) it is  possible to compare the experimental
and     theoretical     fluorescence     spectra     directly     (see
Fig.~\ref{fig:afplot}).    The  top   panel  shows   the  experimental
absorption  spectra  for  2-4  unit  chains. The  next  panel  is  the
experimental    emission   spectra.     As    expected,   experimental
absorption-fluorescence spectra are nearly mirror-image profiles; this
indicates  the same nature  of the  absorbing and  emitting electronic
state.  The  last two panels  are theoretical results  calculated with
broadening parameters of 0.2 eV and 0.02 eV, respectively.  The larger
broadening   is   able   to   model   experimentally   measured   line
shapes. Indeed  we observe good overall comparison  between theory and
experiment. In particular, the shoulder on the red-side of all spectra
are well  reproduced.  The  $\Gamma=0.02$ eV broadening  gives spectra
which show  more detail than  experiment and offer the  possibility of
identification and analysis of  the dominant vibrational modes. In the
theoretical absorption plots, three peaks can easily be resolved. They
correspond, from  left to right, to  a 0-0 transition,  a benzene bond
alternating mode, and  a fast triple bond length  oscillation.  In the
ten unit chain,  we can also resolve a peak  from an alternate benzene
bond  mode.  Figure \ref{fig:modes}  shows schematically  the dominant
normal modes of \pa.  The  top shows the low-frequency stretching mode
I of  the entire molecule, which  contribute for the most  part to the
width  of  vibronic peaks.  Its  frequency  and displacement  $\Delta$
depends on the molecular mass.  These quantities decrease and increase
for  large oligomers,  respectively.   The other  three nuclear  modes
(II-IV) coupled  to the  electronic system are  high-energy vibrations
whose frequencies  weakly depend on  the chain length.  The  first two
(II and III) are bond  alternating modes within the benzene rings that
only become resolved  for the longer chain lengths.   The last mode IV
is  the stretching  of  the triple-bond.   Relative displacements  are
shown for  the individual  modes. The left  value is for  the shortest
chain and the right is for the ten unit chain.  We see that the higher
energy  modes  become  less   dominant  as  the  polymer's  length  is
increased.  The strength  of  the low  energy  stretching mode  nearly
doubles in the long chain limit relative to the short chain limit, and
the  sub-dominant  alternating  modes   such  as  the  second  benzene
alternating mode become resolved in the longer chain length limit.

\begin{figure}[ht]
  \centering
  \includegraphics[width=1.0\textwidth]{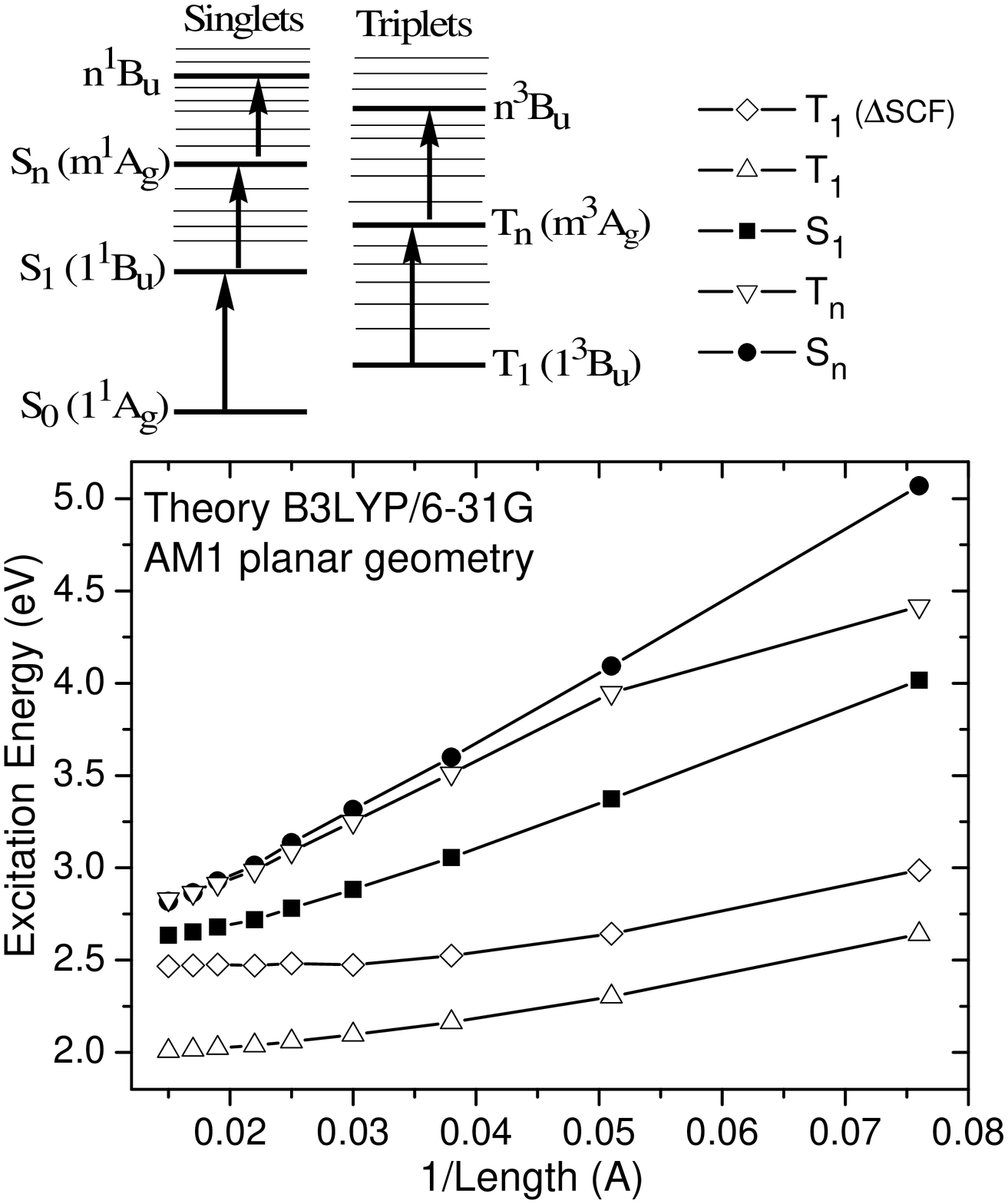}
 \vspace{-0.7in}
\caption{\label{fig:states} Top: Typical electronic structure of optically
active states in singlet and triplet manifolds for conjugated polymers.
Bottom: Size-scaling of excitation energies as a
function of the reciprocal conjugation length for \pa oligomers.}
\end{figure}

Figure \ref{fig:states} shows the size-scaling of calculated low-lying
excited states as a function of reciprocal conjugation length. The top
panel displays a typical structure of singlet and triplet manifolds in
conjugated  polymers  where  the  ladder  of  well-defined  states  is
optically  coupled.  Nearly  linear  scaling  for  all  states  allows
extrapolation to the saturated values  in the infinite chain limit. As
expected,  the singlet-excitation energies  ($S_1$) become  lower with
increasing   polymer  length.   The   long-chain  limit   is  achieved
approximately (within  0.2 eV) by  six units.  The  optically inactive
first  triplet excited  state ($T_1$)  has been  calculated  using two
levels  of  theory:  the   TD-DFT  approach  for  the  lowest  triplet
excitation  and  change  in  the self-consistent  field  ($\Delta$SCF)
method.    The   latter  is   the   energy   difference  between   the
self-consistent  ground states  calculated with  enforced  singlet and
triplet spin multiplicities.  We observe negligible spin contamination
in  the unrestricted  approach used  to calculate  the  triplet ground
state.   In general,  $\Delta$SCF is  considered  to be  a stable  and
reliable approach  for evaluating the  first triplet state  energy. In
contrast,  TD-DFT energies of  triplet states  strongly depend  on the
amount of  the HF exchange  present in the functional.   Even negative
energies  of  triplet  states  may  be observed  in  the  TD-HF  limit
(so-called triplet  instability).  In  our case (B3LYP  functional) we
found 2 and 2.4 eV saturation limits for the energy of the $T_1$ state
of  \pa in  TD-DFT  and $\Delta$SCF  approaches, respectively.  Higher
lying T$_n$  and S$_n$  states correspond to  de-localized excitations
where the electron become well separated from the hole upon absorption
of  a quantum of  light \cite{BredasJL:Excesc}.  Even though  the spin
state matters  for small molecules,  T$_n$ and S$_n$  correctly become
degenerate in  the long chain limit,  saturating to 2.8  eV limit.  In
the  alternating geometry  case, the  electron is  expected  to remain
localized.  All excitations for the alternating geometry are about 0.5
eV  greater than for  the planar  geometry because  of the  more local
nature of the excitation.

\begin{figure}[ht]
  \centering
  \includegraphics[width=1.05\textwidth]{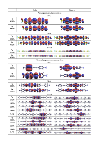}
 \vspace{-0.7in}
\caption{\label{fig:ntos} Selected  transition orbitals for  2, 4, and
10-unit  \pa  chains  in   the  planar  and  alternating  ground-state
geometries. These are calculated  at the B3LYP/6-31G level. The NH$_2$
end is to the right.}
\end{figure}

In  order to  characterize  the typical  low-lying  ($T_1$ and  $S_1$)
excitations  we  further   performed  a  transition  orbital  analysis
\cite{MartinRL:Natto}.   The  top  of  Fig.~\ref{fig:ntos}  shows  the
transition orbitals  for 2, 4,  and 10 unit  chains of planar  \pa. As
expected,  all these states  represent de-localized  transitions which
are mostly  $\pi-\pi^*$ nature. Note  that for the longest  (ten unit)
chains, the  triplets are more  localized than the singlets.   For the
singlet  state, one  pair of  orbitals dominates.   Both the  hole and
particle are de-localized along  the backbone of the polymer; however,
the  hole  is more  de-localized.   Triplet  states  adopt more  of  a
multi-configurational  character with  increasing chain  length, which
reflects  their  smaller exciton  size  compared  to  that of  singlet
states.   The bottom  of Fig.~\ref{fig:ntos}  displays  the transition
orbitals for  the alternating geometry.  These plots  clearly show the
breakage  of   conjugation  due  to  torsion.   All  excitations  have
multi-configurational character as a superposition of orbitals with an
electron-hole pair  residing on short chain segments  (e.g., see $T_1$
for  N=10). Ultimately  such excitations  in long  chains can  be well
treated in the Frenkel exciton limit.

\section{Conclusion}
\label{sec.conclusion}

The  extent  of  electronic   de-localization  and  the  existence  of
appropriate  excited state  energetics have  serious  implications for
using conjugated  polymers as  a constituent in  artificial functional
nano-assemblies.  Our  study confirms  that torsional disorder  of the
molecular  geometry  caused   by  dielectric  environment  or  thermal
fluctuation  is  an  important  factor  affecting  the  excited  state
structure of  \pa. We find  that DFT can give  quantitatively accurate
results  when compared  to  experiment.  Subsequently,  the nature  of
calculated electronic states can be analyzed in the real-space using a
transition  orbital  decomposition.  The  good  agreement between  the
experimental   and  theoretical   fluorescence   spectral  line-shapes
indicate that we can use theory to understand the underlying molecular
morphology   and  to   identify  and   analyze   Franck-Condon  active
vibrational modes.   By comparing our theoretical  calculations to the
experimental results, we find  an effective average geometry and argue
that the conjugation length  is drastically reduced by rotations about
the triple-bonds.

Section \ref{sec.results}  contains only a fraction of  the results we
obtained using combination of  different methods.  Overall, we observe
two  fundamental  trends:  I)  The bond-length  alternation  parameter
reduces with decreasing a fraction  of exact Fock-like exchange in the
DFT   functional  in  the   course  of   the  ground   state  geometry
optimization. This results in  the overall red-shift of the excitation
energies.  II) An  increase of  a fraction  of exact  exchange  in the
functional when  computing TD-DFT excited  states results in  the blue
(red) shift  of singlet  (triplet) state energies.   Consequently, any
combination  of  physically justified  methods  and appropriate  model
chemistries will result  in the same trends and  conclusions with some
variation of calculated spectroscopic variables.

\begin{acknowledgments}
The  research at  LANL is  supported by  Center for  Nonlinear Studies
(CNLS), the  LANL LDRD program,  and the office  of Science of  the US
Department of Energy. This support is gratefully acknowledged.
\end{acknowledgments}


\clearpage
\newpage

\end{document}